\documentclass{PoS}

\title{$t\bar{t}$ Spin Correlations at D0 }

\ShortTitle{$t\bar{t}$ Spin Correlations at D0}

\author{\speaker{Yvonne Peters}\thanks{On behalf of the D0 collaboration.}\\
        University of G\"ottingen and DESY, Germany\\
        E-mail: \email{reinhild.peters@cern.ch}}

\abstract{The heaviest known elementary particle today, the top quark,
  has been discovered in 1995 by the CDF and D0 collaborations at the
  Tevatron collider at Fermilab. Its high mass and short lifetime,
  shorter than the timescale for hadronization, makes the top quark
  a special particle to study. Due to the short lifetime, the top
  quark's 
spin information is preserved in the decay products. In this article
we discuss the studies of $t\bar{t}$ spin correlations at D0, 
testing the full chain from production to decay. In particular, we
present a measurement using angular information and an analysis using
a matrix-element based technique. The application of the
matrix-element based technique to the $t\bar{t}$ dilepton and
semileponic final state resulted in the first evidence for
non-vanishing $t\bar{t}$ spin correlations. }

\FullConference{36th International Conference on High Energy Physics,\\
		July 4-11, 2012\\
		Melbourne, Australia}

\begin{document}

\section{Introduction}

Discovered in 1995 by the CDF and D0 collaborations at the Tevatron
proton antiproton ($p\bar{p}$)
collider at Fermilab, the top quark~\cite{cdfobs,d0obs} is the
heaviest known elementary particle today. The top quark mass is
measured to be $m_t = 173.18 \pm 0.94$~GeV~\cite{topmass}. 
The lifetime of the top quark is shorter than the time scale for
hadronization, therefore it is the only quark that can be studied as
bare quark. 
Due to the short lifetime, the spin information of the top quark is
preserved in its decay products. While $t\bar{t}$ pairs are produced
unpolarized at the Tevatron, 
the correlation of the spin orientation of the top and the anti-top
quark can be studied. 
By investigating the $t\bar{t}$ spin correlations, we can study the
full chain from production to decay and thus test the couplings in
production and decay for 
possible new physics that would change the $t\bar{t}$ spin correlation strength.
In the following, two methods of measuring $t\bar{t}$  spin
correlations, performed by the D0 
collaboration using Tevatron Run~II data, are presented. The first method explores angular distributions, while the second uses a matrix element based approach.

\section{$t\bar{t}$  Spin Correlation Measurement using Angular Distributions}

Despite the unpolarized production of top quark pairs at hadron
colliders, the spins of the top and anti-top quark are expected to be
correlated. 
Information on spin correlations can be extracted from the angular distribution of the final state objects. In particular, the doubly differential cross section $1/\sigma  \times d^2 \sigma/(d\cos \theta_1 d\cos \theta_2)$ can be written as
\begin{equation}
\frac{1}{\sigma} \times  \frac{d^2\sigma}{(d\cos \theta_1 d\cos
  \theta_2)} = \frac{1}{4} \times (1 − A \alpha_1 \alpha_2  \cos \theta_1 \cos \theta_2),
\end{equation}
 where $A$ is the spin correlation strength, $\alpha_1$ ($\alpha_2$) is the spin analysing power of the
final state fermion from the $W^{+}$ ($W^{−}$) boson or top (anti-top) quark decay, and $\theta_1$ ($\theta_2$) is the angle of the 
 fermion in the top (antitop) quark rest frame with respect
to a quantization axis. Several choices of the quantization axis are
common: the helicity axis, where the reference axis is the flight
direction of the top (anti-top) quark in the $t\bar{t}$  rest frame,
the beam axis, where the quantization axis is the beam direction, and the
off-diagonal basis, which yields the helicity axis for ultra-high
energy and the beam axis at threshold. The standard model (SM) prediction of $C =
A\alpha_1 \alpha_2$ depends on the quantization axis and the ratio of
the $t\bar{t}$ production modes. 
At the Tevatron $p\bar{p}$ collider with a center of mass energy of 1.96~TeV, the main $t\bar{t}$  production occurs via
quark-antiquark annihilation to about 85\% and only to about 15\% via
gluon-gluon fusion. 
For the measurement at D0 we consider the beam basis, yielding a SM
prediction of $C = 0.78$ at next-to-leading order (NLO) quantum
chromodynamics (QCD)~\cite{bernreuther}. 
Visually, the spin correlation strength can be considered as the
number of events where top and antitop have the same spin direction
minus the number of events with opposite spin direction, 
normalized to the total number of $t\bar{t}$  events. 
In leading order (LO) QCD, the spin analyzing power $\alpha$ is one
for charged leptons and the down-type quarks from the W boson decay,
and smaller for the up-type quark from the W boson decay and the
$b$-quarks from the top decay.
 Due to the experimental challenge to distinguish up-type from
 down-type quarks, it is easiest to use charged leptons to extract spin correlations.
The D0 collaboration has performed a measurement of $C$ by studying
the distribution $\cos \theta_1 \cos \theta_2$ in the dilepton final
state,  where both $W$ bosons from the top and anti-top quark decay
into a charged lepton and the associated neutrinos, using
5.4~fb$^{−1}$ of Run~II data~\cite{d0coscos}. The measurement is based
on the standard $t\bar{t}$ dilepton selection~\cite{d0dilepxsec},
where two high $p_T$ charged leptons ($ee$, $e\mu$ or $\mu\mu$) of
opposite sign, at least two high-$p_T$ jets and large missing
transverse energy are required.
The main background in this final state arises from $Z$+jets
production, and smaller contributions form diboson production and
instrumental background arising from jets faking a charged lepton. 

In order to calculate $\theta_1$ and $\theta_2$, the reconstruction of
the full $t\bar{t}$ system is required. 
We use the neutrino weighting technique, as developed for precision
top mass measurements~\cite{nuwtopmass}, for this purpose. Neutrino
weighting works as follows: The total
dilepton final state is specified by eighteen components of momentum
from the two charged leptons, two neutrinos and two $b$-jets, of which
only twelve can be measured from the observed jets and charged
leptons. Four additional constraints are provided when requiring that
the invariant mass of a lepton-neutrino pair yields the known $W$
boson mass, and the $W$ boson and $b$-jet combinations yield the top
quark mass. The two additional quantitites that need to be specified
to reconstruct the full event kinematics are extracted by sampling
the pseudo-rapidity distributions of the two neutrinos, providing up
to two solutions for each neutrino transverse momenta. For each
solution a weight is assigned by comparing the measured
value of the missing transverse energy to the calculated missing
transverse energy in the reconstructed event. The resolution of the
$x$ and $y$ components of the missing transverse energy are taken into
account in the weight. Due to the possible jet assignments to the top
quarks, in total eight solutions per event are possible. Detector
resolutions are included in the neutrino weigthing procedure by
smearing the measured lepton and jet momenta according to their
resolution, and by
repeating the calculation for a large number of random choices.

\begin{figure*}[t]
\centering
\includegraphics[width=75mm]{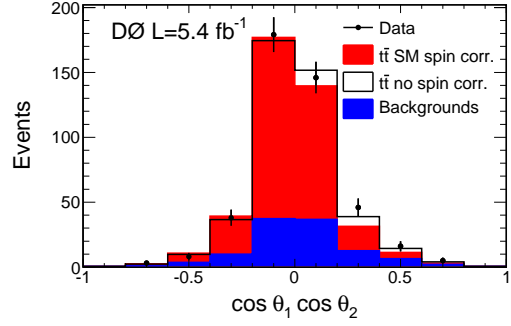}
\caption{The distribution in $\cos \theta_1 \cos \theta_2$ for the
combined  dilepton channel. The expectation of the summed $t\bar{t}$ signal, including
NLO QCD spin correlation (C = 0.777) (red) and all backgrounds
(blue) is shown and are compared to data. The open histogram shows the
$t\bar{t}$ prediction without spin correlation (C = 0)~\cite{d0coscos}. } \label{nuwspin}
\end{figure*}

The extraction of $C$ from $\cos \theta_1 \cos \theta_2$ is performed
by generating a sample including spin correlations at the SM value, 
and a sample neglecting spin correlations ($C = 0$) with the NLO Monte
Carlo (MC) generator MC@NLO~\cite{mcnlo}, 
and building templates in $\cos \theta_1 \cos \theta_2$ for both
$t\bar{t}$ samples and the background, which are fitted to the data. 
We extract $C$ in the beam basis as $C = 0.10 \pm 0.45$~(stat + syst),
in agreement with SM predictions. Systematic uncertainties 
are included as nuisance parameters in the maximum likelihood fit, and
the $t\bar{t}$  cross section is foated freely in order to reduce the
sensitivity to normalization effects. Figure~\ref{nuwspin} shows the
comparison of the predictions with and without $t\bar{t}$ spin
correlations and the data in the combined dilepton final state ($ee$,
$e\mu$ and $\mu\mu$ final states combined).
The measurement is dominated by the statistical uncertainty.
The CDF collaboration has measured $t\bar{t}$ spin correlation using
angular distributions in the dilepton and lepton plus jets final
states~\cite{cdfspin}. 
These measurements also show good agreement with the SM prediction.

\section{$t\bar{t}$ Spin Correlation Measurement using a Matrix-Element based Approach}
The measurement of $t\bar{t}$ spin correlations using angular
distributions is so far limited by the statistical uncertainty. 
Comparing different approaches for the measurement of the top quark
mass, the most precise method is the Matrix Element (ME) method, 
where the full event information is explored. 
The D0 collaboration
explored the application of a ME-based approach for the first time to the measurement of
$t\bar{t}$ spin correlations. 
We test two hypotheses against each other, in particular the
hypothesis of having SM spin correlations ($H = c$) versus the
hypothesis of no spin correlation ($H = u$). 
Per-event signal probabilities $P_{sig}(H)$ are calculated using matrix
elements that include spin correlations or do not include spin
correlations. For hypothesis $H = c$ we use the ME for the
full process $q\bar{q} \rightarrow t\bar{t} \rightarrow W^{+}b W^{-}b
\rightarrow \ell^{+} \nu_{\ell} b \ell^{-} \nu_{\ell} \bar{b}$, averaged
over the initial quarks' color and spin and summed
over the final colors and spins, while for the  hypothesis $H = u$,
we use the ME of the same process, but  neglecting the spin
correlation between production and decay~\cite{parke}. 
We can write $P_{sig}$ as function of the hypotheses, as 
\begin{equation}
P_{sig}(x;H) = \frac{1}{\sigma_{\rm obs}} \int {\rm d}q_1 {\rm
    d}q_2~f_{\rm PDF}(q_1) f_{\rm PDF}(q_2) \frac{(2 \pi)^{4}
    \left|M(y,H)\right|^{2}}{q_1 q_2 s} {\rm d}\Phi_{6} W(x,y) \ ,
\end{equation}
with $\sigma_{\rm obs}$ being the LO $q\bar{q} \rightarrow t\bar{t}$ production cross section
including selection effciency and acceptance effects,
$q_1$ and $q_2$ denoting the fraction of the proton and antiproton
momentum carried by the partons, $f_{\rm PDF}$ representing
the parton distribution functions, $s$ the square of
the center of mass energy of the colliding $p\bar{p}$ system, and
$ {\rm d}\Phi_{6}$ the infinitesimal volume element of the 6-body phase
space. Detector resolution effects are taken into account
by introducing transfer functions $W(x, y)$, that describe
the probability of a partonic final state $y$ to be measured
as $x = (\tilde{p}_1,\dots,\tilde{p}_n)$, where $\tilde{p}_i$  denote the measured four-momenta
of the final state leptons and jets.

\begin{figure*}[t]
\centering
\includegraphics[width=75mm]{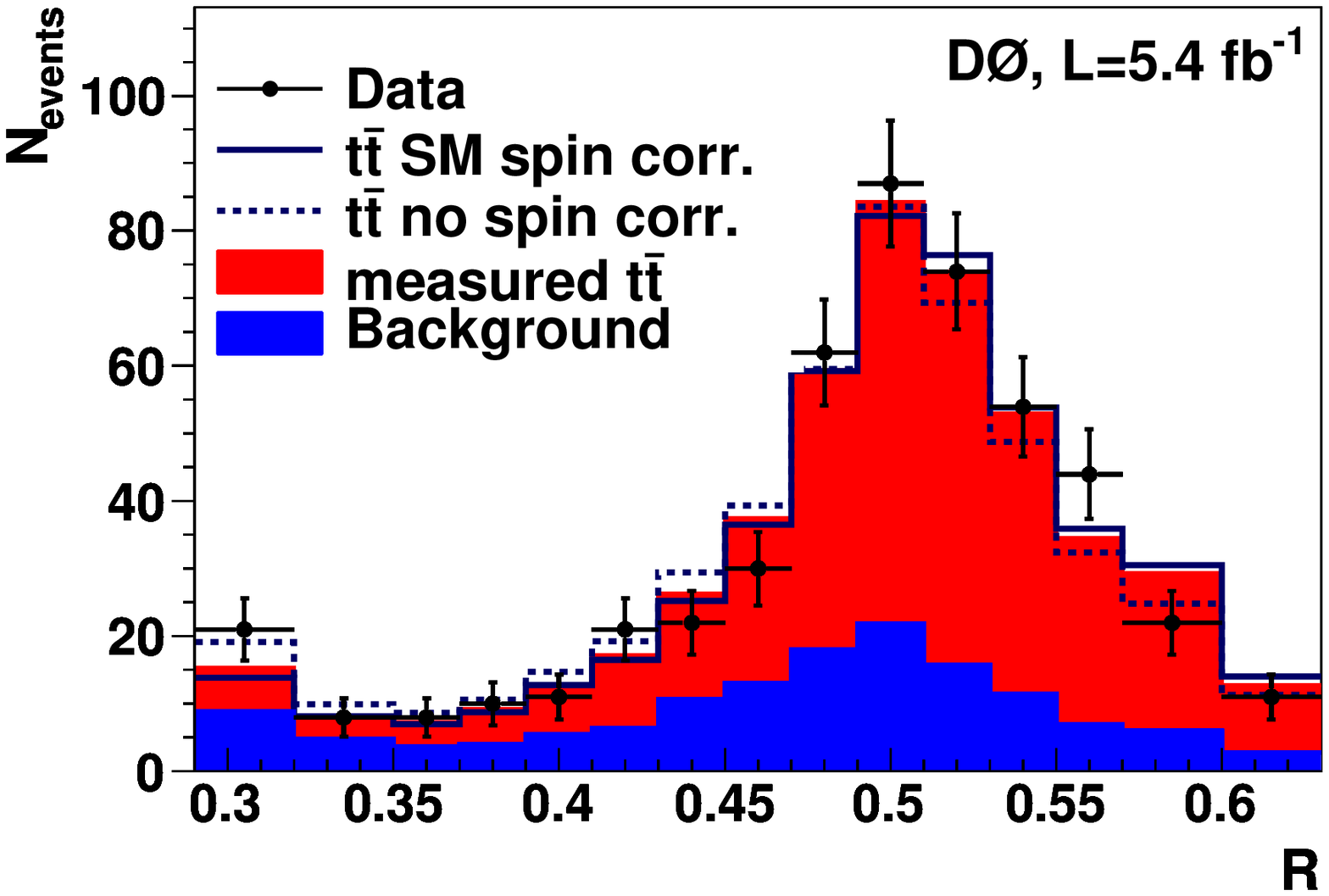}
\includegraphics[width=75mm]{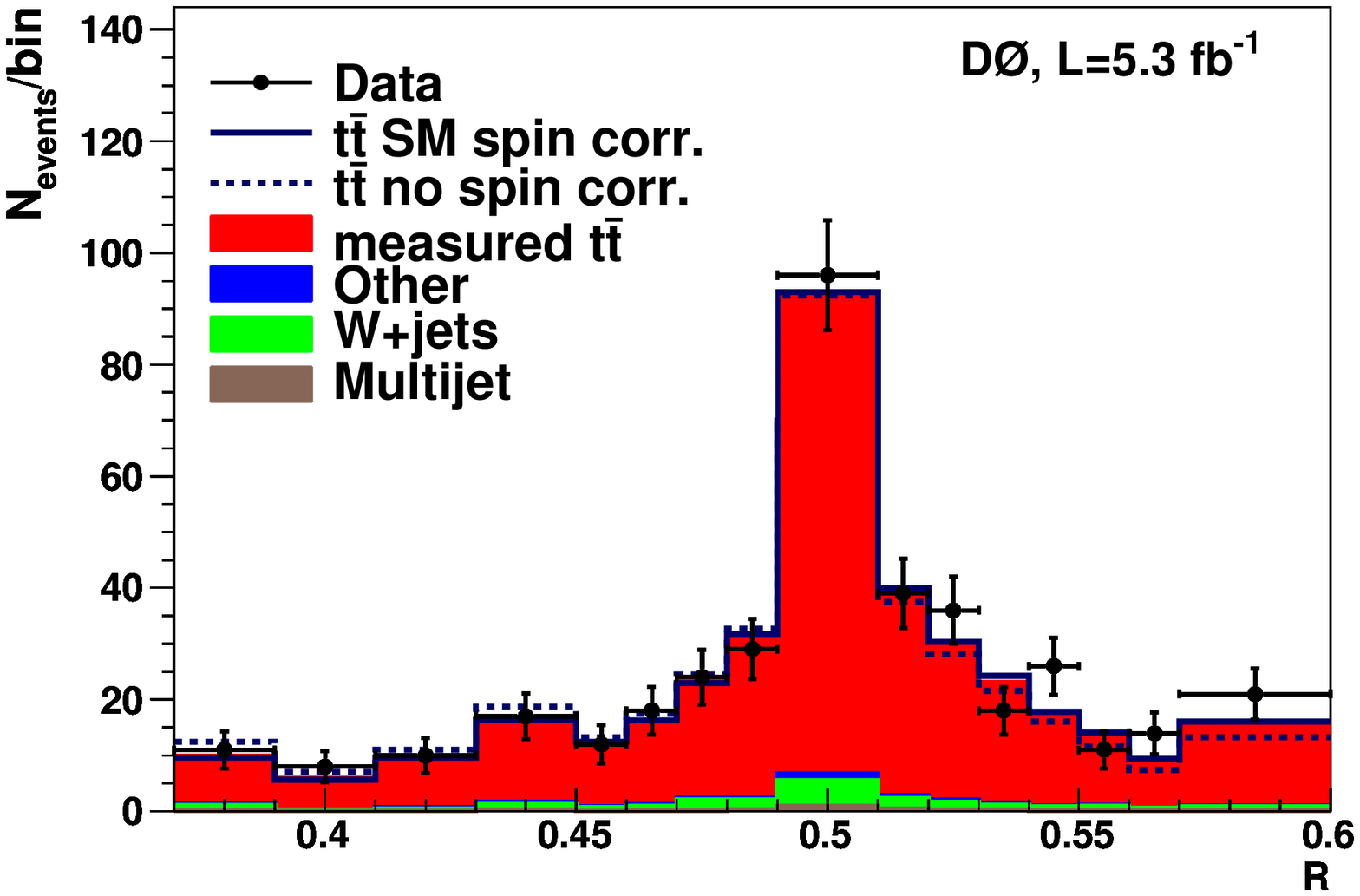}
\caption{ The distribution of the discriminant $R$ for the combined
  dilepton (left) and lepton plus jets (right) final state. The
  expectation for $t\bar{t}$ signal and all backgrounds is shown with
  SM spin correlation  (full line) and without spin correlation (dashed line)~\cite{d0mespin,d0mespinljets}. } \label{mespin}
\end{figure*}

These signal probabilities are then translated into a discriminant~\cite{schulze}:
\begin{equation}
R = \frac{ P_{sig}(H = c) } {P_{sig}(H = c) + P_{sig}(H = u)}.
\end{equation}
Using the same D0 dataset of 5.4~fb$^{-1}$ of dilepton events as for
the measurement with angular distributions, 
a maximum likelihood fit of templates of $R$ has been
performed. Similar to the $t\bar{t}$ spin correlation measurement
using angular distributions, we float the $t\bar{t}$ cross section
freely to reduce the effect from normalization uncertainties on the
measured $t\bar{t}$ spin correlations. 
Samples with different spin correlation content (SM value
and no spin correlations) 
have  been generated using MC@NLO MC, and we use the same samples as
for the measurement using angular distributions. The ME-based approach yields an
improvement of 30\% in sensitivity compared to the analysis using
angular distributions, resulting in $C = 0.57 \pm 0.31$~(stat +
syst)~\cite{d0mespin}. The result is dominated by statistical
uncertainties. 
Figure~\ref{mespin}~(left) shows the comparison of the expected
distributions of the discriminant $R$ for SM
spin correlation and no spin correlation and the data.

While the dilepton final state is the easiest to perform spin
correlation measurements, the D0 collaboration extended the ME-based
measurement to the lepton plus jets final state. 
The selection of semileptonic $t\bar{t}$  events is based on the
$t\bar{t}$ cross section measurement using 5.3~fb$^{−1}$ of
data~\cite{d0ljets}. 
We restrict the sample to events with at least four jets, of which at
least two have to be identified as $b$-jets,  using a neural
network based $b$-tagger that combines variables characterizing the
properties of secondary vertices and tracks displaced with respect to
the primary interaction vertex~\cite{btag}. 
In order to get the right
assignment of final state objects to the top and anti-top, 
four permutations of jets are included: two corresponding to the
choice of which $b$-jet corresponds to the top and anti-top quark, and two
corresponding to 
the assignment of one of the non-$b$-jets to the down-type quark from
the $W$ boson decay. 
To optimize the measurement, we then split the events into four regions
with higher and lower sensitivity. In particular, we distinguish
events according to whether 
exactly four or more than four jets are present, and whether the
invariant mass of the two non-$b$-jets is close or far away from the
known $W$ boson mass. 
The first split is motivated by the fact that for more than four jets,
it is more likely to include wrong jet permutations,
 while the second split is motivated due to a higher probability of
 having picked the wrong jet pair if the invariant mass is far from
 the $W$ boson mass.
 Even though the complication of not knowing the down-type jet reduces
 the sensitivity of the measurement in the lepton plus jets final
 state by about half, the larger dataset,
 about twice as high as in the dilepton final state, yields a
 sensitivity to spin correlations in the lepton plus jets final sate
 similar to the one in the dilepton final
 state. Figure~\ref{mespin}~(right) shows the expectation of signal
 and background for SM $t\bar{t}$ spin correlations and no spin
 correlations compared to the data. 
For the combined fit in the dilepton and lepton plus jets channel, we
measure $C = 0.66 \pm 0.23$~(stat + syst)~\cite{d0mespinljets}, 
which provides the first evidence for non-vanishing $t\bar{t}$ spin
correlations. 

All measurements of $t\bar{t}$ spin correlations are in agreement with
the NLO SM prediction. Independent of the method, the uncertainties of
the results are all dominated by the statistical uncertainty. So far
only half of the full Tevatron data sample has been analysed (5.4 and
5.3~fb$^{-1}$ respectively), and at least a factor of $\sqrt{2}$ of
improvement on the uncertainty can be expected for the final
$t\bar{t}$ spin correlation measurement from D0. Including
improvements on the methods the uncertainty should reduce even
further.

\section{Conclusion and Outlook}
The measurement of $t\bar{t}$ spin correlations provides a test of new
physics in the full chain from production to decay. Only recently, the Tevatron data samples
became large enough to extract sensitive $t\bar{t}$ spin correlation
measurements. 
Several approaches have been explored to measure the spin correlations
strength, in particular a template method using angular distributions,
and a new matrix-element based approach. The application of the latter
to dilepton and lepton plus jets $t\bar{t}$ final states resulted in the first evidence for non-vanishing $t\bar{t}$
spin correlations. 
As the results from Tevatron and LHC are complementary due to
different $t\bar{t}$ production modes dominating, the exploration of
$t\bar{t}$ spin correlation provides one of Tevatron's legacies. 
An important remaining achievement is the exploration of the full Tevatron dataset.

\section*{Acknowledgements}
I thank my collaborators from D0 for their help in preparing the
presentation and this article. I also thank the staffs at Fermilab and
collaborating institutions, 
and acknowledge the support from the Helmholtz association.

\end{document}